\documentclass[prl,twocolumn,groupedaddress]{revtex4-1}

\usepackage{amsmath}
\usepackage{graphicx}
\pdfoutput=1
\usepackage{hyperref}
\hypersetup{colorlinks=true,
			linkcolor=blue,
			citecolor=blue,
			urlcolor=blue,
			pdfborderstyle={/S/U/W1}}

\begin{document}



\title{Importance of orbital fluctuations for the magnetic dynamics in 
heavy-fermion compound SmB$_6$}

\author{Christopher N. Singh}
\email[email at: ]{csingh5@bighamton.edu}
\homepage[visit webpage at: ]{http://www.linkedin.com/in/csingh5binghamton/}

\author{Wei-Cheng Lee}
\email[email at: ]{wlee@binghamton.edu}
\homepage[visit webpage at: ]{http://bingweb.binghamton.edu/~wlee/}

\affiliation{Department of Physics, Applied Physics, and Astronomy, 
Binghamton University, Binghamton, New York 13902, USA}


\date{\today}

\begin{abstract} 
The emergent dynamical processes associated with magnetic excitations in 
heavy-fermion SmB$_6$ are investigated. By imposing multiorbital interactions on a
first-principles model, we find the interplay between spin and orbital
fluctuations in the $f$ manifold is highly sensitive to local correlations. The
magnetic phase diagram constructed at zero temperature reveals quantum critical
features with the existence of several competing phases.  Within the random
phase approximation, we perform a comprehensive study of the spin-spin
correlation function, and our results agree with neutron scattering experiments.
Spectral weight analysis shows the low energy spin excitations are selectively
accompanied by orbital fluctuations, indicating a non-trivial entanglement
between the spin and orbital degree of freedom driven by relativistic couplings.
\end{abstract}


\maketitle

\textit{Introduction}.
While the possibility that SmB$_6$ is topologically nontrivial has driven many
recent efforts \cite{menthe1969, kouwenhoven2001, hatnean2013, dzero2010,
wolgast2013, kim2013, kim2014, neupane2013, xu2014, wakeham2016, li2014,
tan2015}, an equally relevant aspect of this material has been brought to light
through the lens of inelastic magnetic neutron scattering (INS). Specifically,
the temperature activated \cite{nyhus1997} dynamical magnetic response
signatures observed deep within the insulating state are not traditional
magnons, show a high degree of momentum space anisotropy, and have been
attributed to correlation driven exciton \cite{alekseev1995, fuhrman2014,
fuhrman2015} modes. The narrow gap, strong Coulomb interaction, and residual
specific heat give exciton-type modes considerable plausibility in the context
of this system \cite{knolle2017}, and identifying the extent to which these
excitations contribute to the low-energy transport properties, as well as the
interplay between correlation and topology is crucial in understanding SmB$_6$.
In fact, it is well known in the heavy fermions that the Coulomb interaction,
lattice geometry, and spin orientation are essential in spawning exotic
phenomena; however, fair treatment of the multiorbital nature is often hindered
by an exponential growth in complexity. It is precisely this interplay of
competing energy scales and many degrees of freedom that invoke the striking
electronic properties, yet in spite of this, a multiorbital, first principles
study of the magnetic dynamics in SmB$_6$ is still lacking.

We address this gap with a realistic model based on complementing density
functional theory (DFT) with the generalized random phase approximation (GRPA).
This is achieved by projecting the relativistic eigenstates of the Kohn-Sham
equations onto Wannier functions, and imposing the multiorbital
Hubbard-Kanamori interaction onto these maximally localized orbitals. This
approach treats the spin-orbit coupling, multiorbital Coulomb interaction, and
band-structure effects on equal footing. Quantum critical features are found at
zero temperature with several nearby magnetic phases. In the normal state at
finite temperature, the low-energy spin excitations are shown to be tightly
coupled to orbital exchange processes through the large spin-orbit coupling, and
a number of important features observed in the INS experiments naturally emerge
with this approach.

\textit{Model}.
Motivated to capture hybridization effects between localized Sm 4$f$ moments,
and itinerant Sm 5$d$ states, we employ a relativistic multiorbital Hamiltonian
as
\begin{equation} 
	H = H_t + H_{int} 
\end{equation} 
where $H_t$ is given by 
\begin{equation}
	H_t = \frac{1}{2}\sum_{ij\alpha \beta \sigma} (t_{ij}^{\alpha \beta} - 2
		\mu \delta_{ij} \delta_{\alpha \beta}) c^{\dagger}_{i \alpha \sigma}
		c_{j \beta \sigma}
\end{equation}
Here the fermion operators create (destroy) particles at site $i$ $(j)$, with
orbital character $\alpha$ $(\beta)$ and spin $\sigma$. Symmetry considerations
and the spin-orbit interaction dictate the Wannier basis is chosen as spinors of
the Sm $d$-$eg$ states and the full Sm $f$ level multiplet \cite{chang2015}.  In
this way, contained within $H_t$ is the fully relativistic \textit{ab initio} information
pertaining to the entirety of the $d$-$f$ hybridization, as well as $f$ level
character in the vicinity of the Fermi energy. This approach has the advantage
of treating the $f$ manifold relativistically in contrast to previous studies
\cite{fuhrman2014}, and is known to be sufficient in producing the hybridization
gap \cite{yanase1992}. 
\begin{equation} 
	\begin{split} 
		H_{int} = \hspace{1mm}  & \frac{U}{2}\sum_{i \alpha \sigma}
		n_{i \alpha \sigma} n_{i \alpha \sigma^{\prime}} \\ 
		&+ \sum_{i,\alpha<\beta,\sigma} \big\{ (U-2J) n_{i \alpha \sigma}
		n_{i \beta \sigma^{\prime}} + (U - 3J)n_{i \alpha \sigma} 
		n_{i \beta \sigma} \\
		&+ J(c^{\dagger}_{i \alpha \sigma} c_{i \beta \sigma} 
		c^{\dagger}_{i \beta \sigma^{\prime}} c_{i \alpha \sigma^{\prime}} -
		c^{\dagger}_{i \alpha \sigma} c_{i \beta \sigma}
		c^{\dagger}_{i \alpha \sigma^{\prime}} c_{i \beta \sigma^{\prime}}) \big\} 
	\end{split} 
\end{equation} 
$H_{int}$ is the centrosymmetric representation \footnote{The centrosymmetric
approximation is used for simplicity, but some recent evidence \cite{xiang2017}
suggests that a more detailed treatment of the inter-orbital interaction may be
pertinent.} of the multiorbital Hubbard-Kanamori interaction \cite{kanamori}
that is treated at the mean field level to calculate the magnetic phase diagram,
and at the RPA level to calculate the dynamical spin-susceptibility in the
normal state. $U$ is the intra-orbital repulsion, and $J$ is the Hund's coupling
parameter. The first principles calculations are performed with full potential
linear augmented plane waves plus local orbitals and the local density
approximation implemented in the WIEN2k \cite{blaha2001} ecosystem. The total
energy was converged to 0.1 meV on a 5000 \textit{k}-point grid with an RKmax of 5.
Projection onto Wannier states including fourth nearest neighbors is
accomplished with the Wannier90 package \cite{mostofi2014}, resulting in 40,500
complex hopping parameters. 

\begin{figure}
	\includegraphics[width = \linewidth]{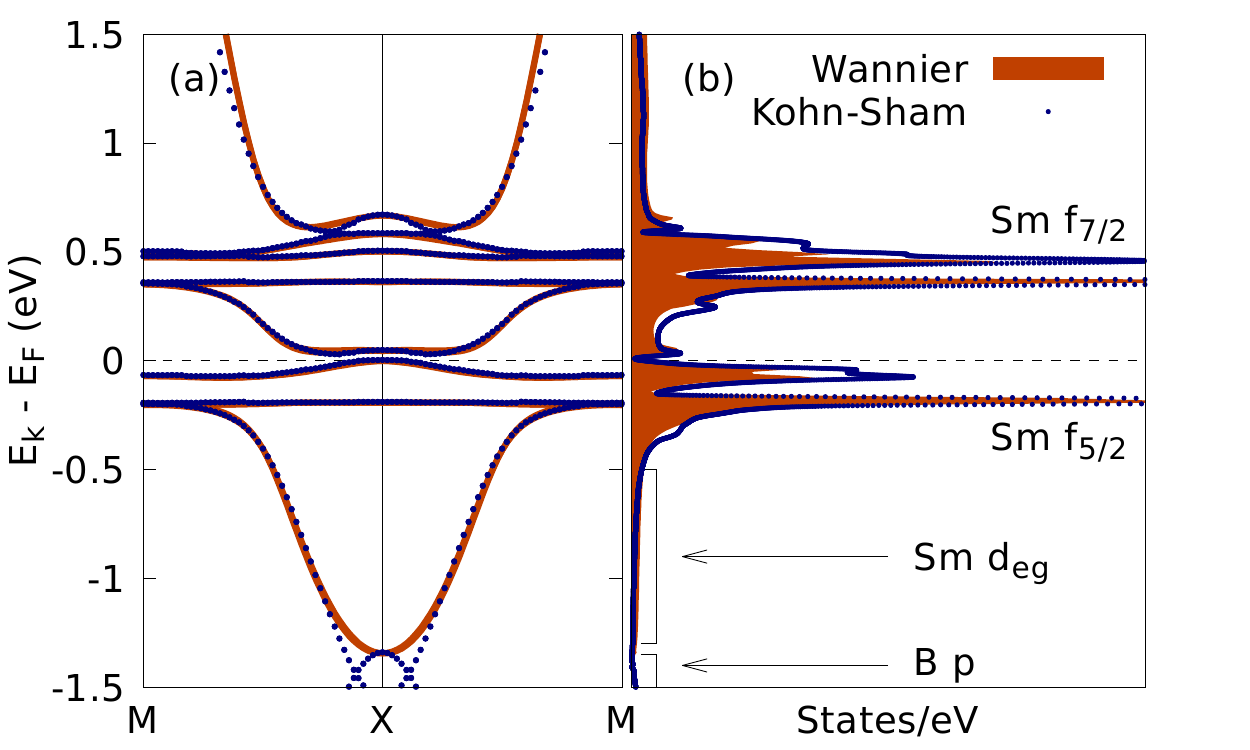}
	\caption{\label{dft} Electronic structure plots contrasting local-density approximation
	 and Wannier projection. (a) Band-structure; (b) density of states.}
\end{figure}

Figure \ref{dft} overlays the Wannier interpolated electronic structure with the
Kohn-Sham result. The density of states shows the sharp Sm $f$ peaks with the
doubly split $J = 5/2$ and triply split $J = 7/2$ multiplets just below and
above the Fermi level respectively \cite{chang2015}. This electronic structure
is representative of $O_h$ point group symmetry in a weak cubic field and strong
spin-orbit coupling scheme \cite{tinkham2003}; in this respect it is
commensurate with the latest tunneling spectra \cite{sun2018}.  The itinerant Sm
$d$-$eg$ bands are seen to hybridize with the localized $f$ manifold developing
a 15 meV direct gap. Excellent agreement is found between the Wannier projection
and Kohn-Sham result in the low energy window $E_f \pm 500$ meV.  Admittedly, a
parity crossing between the hybridized Samarium $4f$ band and the Boron $p$
state at the X point lost in this Wannier projection, likely resulting in a
shift of the Berry phase. However, being interested in excitation effects, this
truncated basis serves as an effective representation of the low energy physics.

\textit{Mean Field Theory} -- 
Decoupling the quartic terms in the interaction is 
accomplished as in Refs. \cite{nomura2000, dagotto2011} with 
\begin{equation}
	\langle c^{\dagger}_{i \alpha \sigma} c_{j \beta\sigma \prime} \rangle = 
	[n_{\alpha} + \frac{\sigma}{2} 
	\text{cos}( \mathbf{q \cdot r_i}) m_{\alpha}] 
	\delta_{ij} 
	\delta_{\alpha \beta}
	\delta_{\sigma \sigma \prime} 
\end{equation} 
This leads to a momentum space mean field Hamiltonian
\begin{equation} 
	\begin{split}
		H^{MF} = 
		H_t &+ \sum_{\mathbf{p} \alpha \sigma} \theta_{\alpha} 
		c^{\dagger}_{\mathbf{p} \alpha \sigma} 
		c_{\mathbf{p} \alpha \sigma} + \zeta \\ 
		&+ \sum_{\mathbf{p} \alpha \sigma} \eta_{\alpha \sigma} 
		(c^{\dagger}_{\mathbf{p} \alpha \sigma} c_{\mathbf{p+q} \alpha \sigma}
		+ h.c) 
	\end{split} 
\end{equation} 
with mean field potentials
\begin{equation} 
	\begin{split} 
		&\theta_{\alpha} = Un_{\alpha} + (2U - 5J) \sum_{\beta \neq \alpha} 
		n_{\beta} \\ &\eta_{\alpha \sigma} =
		-\frac{\sigma}{2} \big( Um_{\alpha} + J\sum_{\beta \neq \alpha} m_{\beta} \big)
	\end{split} 
\end{equation} 
and mean field constant
\begin{equation} 
	\begin{split} 
		\zeta = \frac{J}{2} \sum_{\alpha \neq \beta} m_{\alpha}m_{\beta} 
		&- U \sum_{\alpha}(n^2_{\alpha} - \frac{1}{4}m_{\alpha}^2 ) \\ 
		&- (2U - 5J) \sum_{\alpha \neq \beta} n_{\alpha}n_{\beta} 
	\end{split} 
\end{equation} 
Calculating the phase diagram proceeds by self-consistently
determining the mean field parameters $n_{\alpha}$ and $m_{\alpha} = n_{\alpha
\uparrow} - n_{\alpha \downarrow}$, with convergence characterized by 
$||D|| < 1\times10^{-5}$.
\begin{equation} 
	D = \langle n_{i+1}^{\alpha} - n_{i}^{\alpha} |^{\frown} 
		\langle m_{i+1}^{\alpha} - m_{i}^{\alpha}| 
\end{equation} 
Minimization of the norm of $D$ gives the self-consistent condition,
automatically ensuring a minimum in the free energy \cite{johnson1988}. The
self-consistent process is repeated across different magnetic phases and
ordering wavevectors. We consider a set of 5 phases characterized by 3
antiferromagnetic ordering wavevectors $\mathbf{q}_1 = (\frac{1}{2},0,0)$,
$\mathbf{q}_2 = (\frac{1}{2},\frac{1}{2},0)$, $\mathbf{q}_3 =
(\frac{1}{2},\frac{1}{2},\frac{1}{2})$, and the paramagnetic and ferromagnetic
phases. The total particle number is constrained to the experimental average Sm
valence of 2.54 \cite{utsumi2017} during each of the self-consistency cycles. 

\begin{figure} [ht!]
	\includegraphics[scale = 0.28]{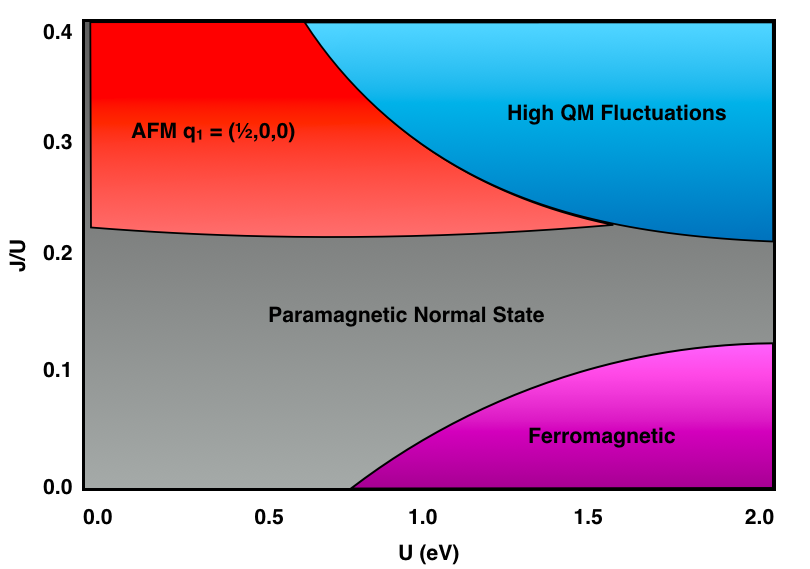}
	\caption{\label{phase} Schematic magnetic phase diagram of SmB$_6$ obtained
	by mean-field treatment of first principles Wannier projection.} 
\end{figure}

Figure \ref{phase} shows the zero temperature magnetic phase diagram in the
plane of the interaction parameters $U$ and $J$. A central feature consistent
with $\mu SR$ experiments \cite{biswas2014} is the large paramagnetic belt found
in the region with moderate correlations where the intra-orbital repulsion is
comparable to the $f$ level bandwidth $W$.  Interestingly, in the regime of
large Hund's coupling $J$ compared to intra-orbital repulsion $U$, $\mathbf{q}_1
= (\frac{1}{2},0,0)$ antiferromagnetic order is found to be the groundstate. We
notice that high pressure experiments \cite{barla2005} have already seen
evidence for this 1-D like antiferromagnetic order, and a recent theoretical
study reported in Ref. \cite{chang2017} has obtained similar results.  The
region of $U > 1$eV, $J/U > 1/5$ shows several phases very close in energy,
suggesting the dominance of quantum fluctuations and highly competing order. For
$U >> J/U$, ferromagnetism is found to be the lowest-energy magnetic phase.  It
is worth mentioning that experimental evidence for ferromagnetic order is not
conclusive. While $\mu SR$ experiments \cite{biswas2014} find no evidence of
long range ferromagnetic order, magnetoresistance experiments
\cite{nakajima2016} are suggestive of ferromagnetic puddling. In short, our
mean-field calculations are commensurate with experiments in suggesting a system
with various competing magnetic orders at zero temperature, implying the
magnetic dynamics are complicated even in the normal state at finite
temperatures.

\textit{Spin susceptibility}.
In order to deepen our understanding of the spin
dynamics in multiorbital spin-orbit coupled systems, we study the magnetic
excitations in the normal state with the following correlation tensor
\begin{equation} 
	\mathring{\chi}^{\gamma	\delta}_{\alpha \alpha^{\prime} \beta \beta^{\prime}}
	(\mathbf{q} , i \omega_n) = 
	\int_0^{\beta} d \tau e^{i \omega_n \tau} \langle T_{\tau}
	m^{\gamma}_{\alpha \alpha^{\prime}}(\mathbf{q}, \tau) m^{\delta}_{\beta \beta^{
	\prime}}(-\mathbf{q}, 0) \rangle 
\end{equation} 
where for example
\begin{equation} 
	m^z_{\alpha \alpha^{\prime}}(\mathbf{q}, \tau) =
	\sum_{\mathbf{p} \sigma} 
	\sigma c^{\dagger}_{\mathbf{p + q} \alpha \sigma} (\tau)
	c_{\mathbf{p} \alpha^{ \prime} \sigma} (\tau) 
\end{equation}
The lower Greek indices represent orbitals and the upper indices represent
magnetization direction components. Evaluation of the correlation tensor follows
textbook procedures \cite{mahan}, and the bare susceptibility can be expressed
by the generalized Lindhard function  
\begin{equation} 
	\begin{split}
		\mathring{\chi}^{zz}_{\bar{\alpha} \bar{\beta}} (\mathbf{q}, \omega) &= 
			\frac{1}{2N} \sum_{\mathbf{p} \sigma} \sigma 
			\Xi_{\bar{\alpha} \bar{\beta}}^{ab \sigma}(\mathbf{p, q}) 
			\Lambda_{ab}(\mathbf{p, q},\omega) \\
		\Xi_{\bar{\alpha} \bar{\beta}}^{ab \sigma}(\mathbf{p, q}) &\equiv
			(U^{\mathbf{p + q}}_{\alpha a \sigma})^* 
			U_{\alpha^{\prime} b \sigma}^{\mathbf{p}} 
			(U^{\mathbf{p}}_{\beta b \sigma} )^* 
			U_{\beta^{\prime} a \sigma}^{\mathbf{p + q}} \\ 
		\Lambda_{ab}(\mathbf{p, q}, \omega) &\equiv
			\frac{n_F(\xi_{\mathbf{p + q} a}) - n_F(\xi_{\mathbf{p} b}) } 
			{\omega + i \eta + \xi_{\mathbf{p} b} - 
			\xi_{\mathbf{p+ q} a}} 
	\end{split}
\end{equation} 
where contravariant indices are eigenbasis indices and are summed over, $\Xi$ is
the orbital projection weight, and $\Lambda$ gives the thermal occupations. 
Here the rank four tensor is operated
as a matrix by defining the sets $\bar{\alpha} = \{\alpha,\alpha^{\prime} \}$
and $\bar{\beta} = \{\beta, \beta^{\prime} \}$. Due to the presence of strong
spin-orbit coupling, the longitudinal ($\chi^{zz}$) and transverse
($\chi^{\pm}$) functions are calculated separately since they could be
different. Within the GRPA, the renormalized correlation functions become
\begin{equation} \label{grpa_eqns}
	\begin{split}
		&\chi^{zz}_{\bar{\alpha} \bar{\beta}} = \chi^1 + \chi^4 - \chi^2 - \chi^3 \\
		&\chi^{\pm}_{\bar{\alpha} \bar{\beta}} = \chi^5
	\end{split}
\end{equation}
The  functions $\chi^{1-5}$, along with the interaction kernel are worked out in
great detail in Refs. \cite{wu2015, mukherjee2016}. The spectral function of
this correlator is directly measured by INS experiments, and what is known from
experiment is the low energy peaks around 14 meV cannot be attributed to phonon,
crystal field, or pure magnon modes \cite{alekseev1995}.

\textit{Discussion}.
To gain insight into the origin of these peaks, we analyze the orbital
components of the spectra around 14 meV for a set of scattering vectors tested
by Ref.  \cite{alekseev1995}. The GRPA calculations were performed on an 8000
$k$-point grid in the full Brillouin zone with a thermal broadening factor fixed
to $\eta = 0.5$ meV.
\begin{figure} 
	\centering
	\includegraphics[width = \linewidth]{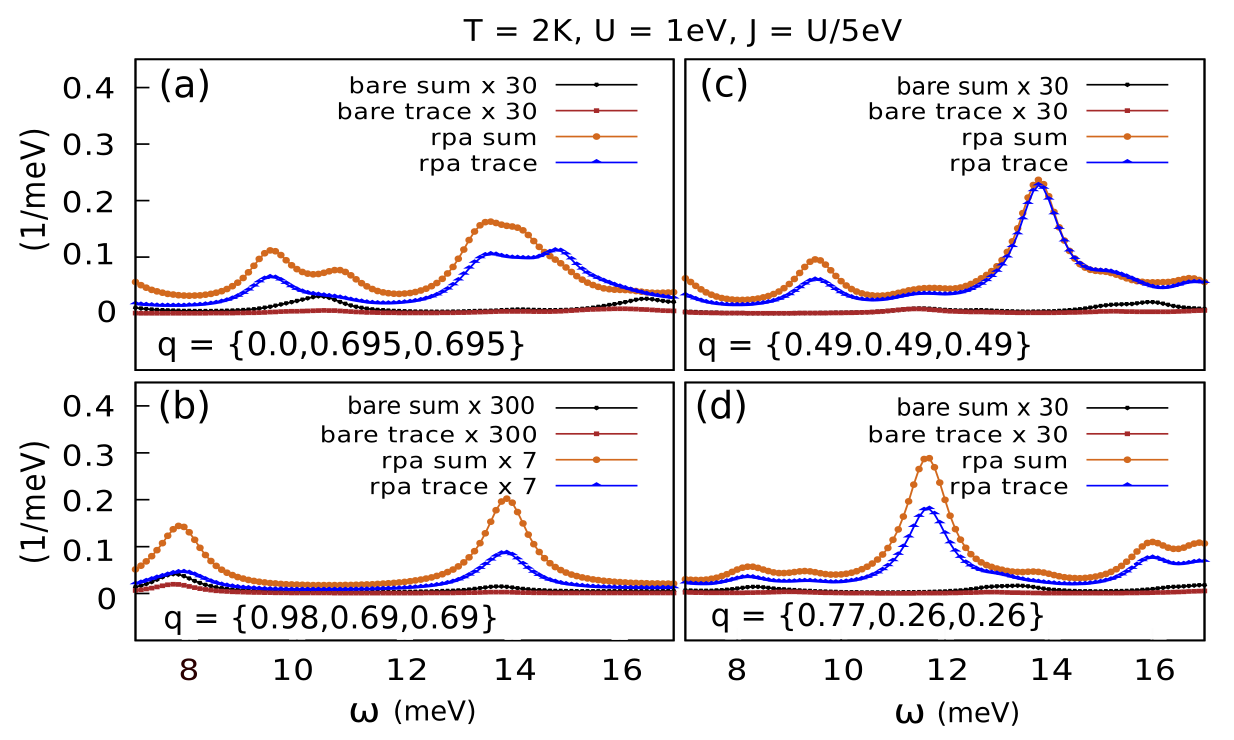}
	\caption{\label{4way} Longitudinal spin response spectrum for selected
		scattering vectors of INS data of Ref. \cite{alekseev1995}. 
		Note the spectra have been scaled
		differently.}
\end{figure}
Table \ref{table} summarizes how the correlation tensor is used to classify
processes depending on initial and final orbital states, and figure \ref{4way}
shows two orbital decompositions of the spectral function extracted via the sum
and the trace of the longitudinal function from Eqn \ref{grpa_eqns}. 
\begin{table}[]
	\caption{\label{table} Sum and trace operations on the correlation tensors are
	used to determine if orbital fluctuations are present in that excitation channel} 
	\begin{ruledtabular} 
		\begin{tabular}{c c} 
		Function & Orbital conservation \\ 
		\hline \hline 
		$\sum_{\bar{\alpha} \bar{\beta}}$( $\chi^{\pm}_{\bar{\alpha} \bar{\beta}}$ ) & No \\ 
		$\sum_{\bar{\alpha} \bar{\beta}}$( $\chi^{zz}_{\bar{\alpha} \bar{\beta}}$ ) & No \\
		tr( $\chi^{\pm}_{\bar{\alpha} \bar{\beta}}$ ) & Yes  \\
		tr( $\chi^{zz}_{\bar{\alpha} \bar{\beta}}$ ) & Yes \\
		\end{tabular} 
	\end{ruledtabular} 
\end{table}

Consider first the bare and the GRPA susceptibilities in figure \ref{4way}$a$.
The bare function shows no signature at 14 meV whereas the GRPA produces peaks
matching the INS data, indicating these modes are a result of electron
correlations instead of wavevector nestings. Furthermore, the difference between
the sum and the trace of the spectral function demonstrates the extent to which
spin excitations at that wavevector have considerable orbital content. This is
readily visible in comparing figure \ref{4way}$a$, \ref{4way}$b$, and
\ref{4way}$d$ to figure \ref{4way}$c$. If the sum and the trace have nearly
identical line-shapes, the corresponding peak is mainly associated with
spin-flip processes within intra-orbital channels. In this case, orbital
fluctuations are not coupled to this spin mode despite the strong spin-orbit
interaction.  On the other hand, if the trace is only a portion of the the sum
around an INS peak, the corresponding peak carries significant weight in the
inter-orbital channel, and orbital fluctuations are strongly entangled with this
spin excitation. 
\begin{figure}[t!]
	\centering
	\includegraphics[width = \linewidth]{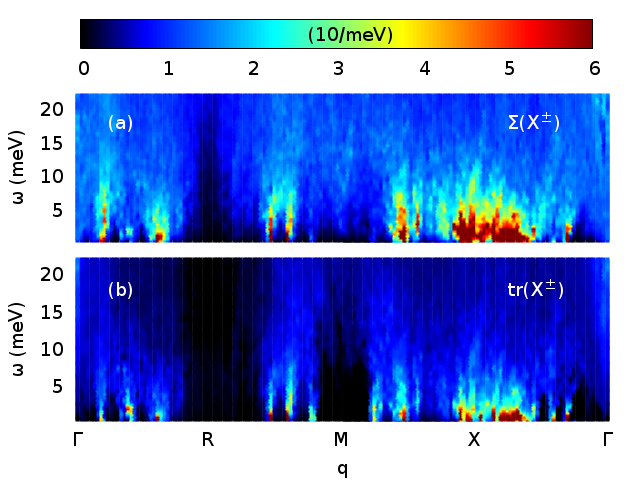}
	\caption{\label{U1J01K3pm} 
		(top) The sum of the transverse function for $T = 1$K, $U = 1$eV, $J = U/10$. 
		(bottom) trace of the transverse function for $T = 1$K, $U = 1$eV, $J = U/10$.}
\end{figure}
Comparing the spin-spin correlation function at all four momenta plotted in
figure \ref{4way}, we find that the magnetic excitations at $q =
(0,0.695,0.695)$, $(0.98,0.69,0.69)$ and $(0.77,0.26,0.26)$ have large
inter-orbital contributions while those at $q = (0.49,0.49,0.49)$ are mainly in
the intra-orbital channels.  This observation indicates that the effects of the
interactions driving the spin collective modes near 14 meV are inhomogeneous
throughout the momentum space, despite the fact that SmB$_6$ has a
centrosymmetric crystal structure and the Sm point group should be at lowest
$D_{4h}$.  This strongly suggests the orbital degree of freedom plays a crucial
role in the collective excitations emerging from electron correlations, and may
lead to the symmetry breaking magnetic response witnessed in Ref.
\cite{xiang2017} for example.

Figure \ref{U1J01K3pm} maps the transverse spin excitation in frequency-momentum
space. The inter- vs intra-orbital channels can be seen to have different
structure and intensity as a function of scattering wavevector.  This reiterates
a strong inhomogeneity in the spin-orbital coupling,  and supports the idea that
the low energy states are selectively susceptible to orbital excitations. The
peak around $\mathbf{q} = $ X observed in INS and in our data can be directly
tied to the phase diagram, as this excitation is associated with 1D AFM order.
The lowest lying excitations exhibiting a reduced dimensionality profile has
ramifications on transport properties as discussed by Ref. \cite{knolle2017},
especially given the centrosymmetry present in the crystal structure and our
interaction kernel. The fact that we find the X point susceptibility peaking
near 4 meV is indicative that in our model, the cost of the 1D AFM excitation is
within 10 meV of experiment. Given the fact we are in a weak coupling regime,
this suggests that even though this is correlation driven physics, the $U
\rightarrow \infty$ limit is not absolutely necessary. This alleviates chemical
potential pinning and integer occupancy constraints imposed by slave
bosonization for example, and is another benefit of this approach to mixed
valent system. To explore this further, the onset of the excitation is studied
as a function of local correlations.

\begin{figure}[t!] 
	\centering
	\includegraphics[width = \linewidth]{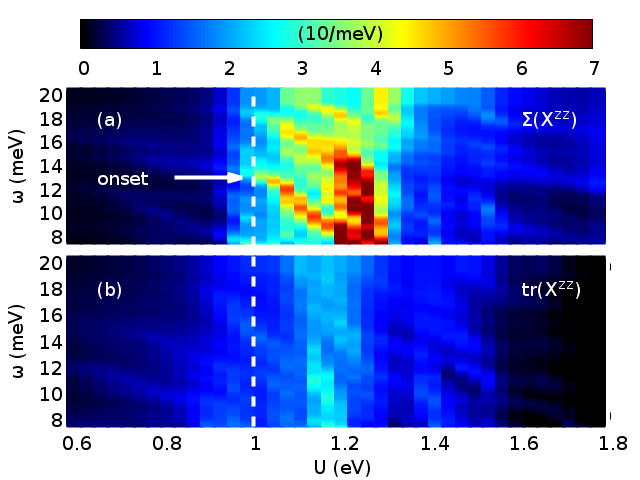}
	\caption{\label{Umap}
		Scattering at q = (0,0.695,0.695) as a function of interaction for $T = 2$K, $J = U/5$.
		(a) The sum of the longitudinal function.
		(b) trace of the longitudinal function. }
\end{figure}

Figure \ref{Umap} shows the longitudinal susceptibility as a function of $U$
with $J = U/5$ for the selected scattering vector $\mathbf{q} = (0, 0.695,
0.695)$. We find that the excitation at 14 meV onsets as $U \approx W \approx
1$eV, and is driven down in energy as a function of $U$. While the result agrees
with the previous study \cite{fuhrman2014, fuhrman2015}, our results further
ascribe significant orbital angular momentum to the spectral weight of the 14
meV mode by comparing the trace and the sum. The trace in figure \ref{Umap}$b$
showing a significantly weaker excitation profile than the sum in \ref{Umap}$a$,
again shows that at this specific wavevector there is significant orbital
character in the excitation. In light of the phase diagram in figure
\ref{phase}, increasing $U$ drives the system into a region of high
fluctuations, reducing the energy cost of instantiating this specific
spin-orbital excitation. The fact that the 14 meV peak arises when $U \approx W$
places a strong constraint on theoretical treatment of correlations in SmB$_6$,
further showing the $U\rightarrow\infty$ limit is an unnecessary assumption if
starting with an accurate electronic dispersion.

\textit{Conclusion}.
We have shown that a first principles model can reproduce
the low energy physics in SmB$_6$, and that momentum dependent entanglement
between the spin and orbital degree of freedom emerges naturally from strong
spin-orbit coupling. The various competing magnetic phases at zero temperature
lead to non-trivial magnetic dynamics in the normal state, and spectral
decomposition of the spin susceptibility exposes the anisotropic orbital
character of the excitations. This first-principles approach clarifies a number
of intriguing features observed in the inelastic neutron scattering measurement.
With the evidence presented here, we propose the orbitally degenerate
non-dispersive $f$ manifold is the perfect environment to harbor \textit{orbital
exciton} modes, a new correlation driven mode carrying exclusively orbital
angular momentum. This conjecture proffers a different physical interpretation when
considering non-trivial topology with a charge-neutral Fermi surface, and
provides a simple mechanism for bulk SmB$_6$ to couple selectively to magnetic
perturbations while simultaneously ignoring the charge sector. This exciton 
form allows an additional pathway for low temperature specific
heat anomalies, and will additionally cause an orbital dichroic signal in
optical probes. Further work to address the role of topology, as well as
quantitative descriptions of these contemporary exciton modes is underway.

\begin{acknowledgments} 
\textit{Acknowledgments}.
This work utilized the Extreme Science and
Engineering Discovery Environment (XSEDE) supported by National Science
Foundation grant number ACI-1548562. It was also supported by a Binghamton
University start up fund.  The authors thank Pegor Aynajian for discussions
regarding INS and Feliciano Giustino for discussions regarding DFT.
\end{acknowledgments}

\end{document}